\documentclass[twocolumn,secnumarabic,amssymb, nobibnotes, aps, prd]{revtex4}
\usepackage{graphicx}

\begin{document}
\title{Dynamic processes in superconductors and the laws of thermodynamics }

\author{A.V. Nikulov}
\email[]{Telephone: +74965244135; fax: +74965244225; e-mail: nikulov@iptm.ru}
\affiliation{Institute of Microelectronics Technology and High Purity Materials, Russian Academy of Sciences, 142432 Chernogolovka, Moscow District, RUSSIA.} 

%nikulov@ipmt-hpm.ac.ru
%\date{}
\begin{abstract} The transition from the superconducting to the normal state in a magnetic field was considered as a irreversible thermodynamic process before 1933 because of Joule heating. But all physicists became to consider this transition as reversible after 1933 because of the obvious contradiction of the Meissner effect with the second law of thermodynamics if this transition is considered as a irreversible process. This radical change of the opinion contradicted logic since the dissipation of the kinetic energy of the surface screening current into Joule heat in the normal state cannot depend on how this current appeared in the superconducting state. The inconsistency of the conventional theory of superconductivity, created in the framework of the equilibrium thermodynamics, with Joule heating, on which Jorge Hirsch draws reader's attention, is a consequence of this history. In order to avoid contradiction with the second law of thermodynamics, physicists postulated in the thirties of the last century that the surface screening current is damped without the generation of Joule heat. This postulate contradicts not only logic and the conventional theory of superconductivity but also experimental results.  
 \end{abstract}

\maketitle

\narrowtext

\section{Introduction}
\label{}
Jorge Hirsch draws reader's attention on very important fundamental problems in the articles published in Physica C \cite{Hirsch2020Physica} and other journals  \cite{Hirsch2020EPL,Hirsch2020ModPhys,Hirsch2020APS}. Hirsch concludes that the conventional theory of superconductivity cannot be correct \cite{Hirsch2020APS} because of their contradiction with the laws of thermodynamics \cite{Hirsch2020Physica,Hirsch2020EPL,Hirsch2020ModPhys}. He does not consider a possibility that equilibrium thermodynamics may not be applicable to the description of superconducting transition and dynamic processes in superconductors. The history of the concepts about superconductivity and results of experimental investigations of dynamic processes in superconducting rings, published in Physica C \cite{Physica2019} and other journals \cite{PLA2012Ex,Letter2007,JETP2007,Letter2003}, give evidence this possibility. 

\section{The superconducting transition in a magnetic field was considered as irreversible before 1933}
Hirsch has concluded that the conventional theory of superconductivity is internally inconsistent \cite{Hirsch2020ModPhys} since on the one hand, this theory is created within the framework of equilibrium reversible thermodynamics, and on the other hand, it predicts Joule heating. Therefore, it is important to recall that the transition from the superconducting state to the normal state in a magnetic field was considered as a non-equilibrium irreversible process before the discovery the Meissner effect in 1933 \cite{Meissner1933}. D. Shoenberg wrote in the book \cite{Shoenberg1952} published in 1952: "{\it At that time} [before 1933], {\it it was assumed that the transition in a magnetic field is substantially irreversible, since the superconductor was considered as a perfect conductor (in the sense which was discussed in Chapter II), in which, when the superconductivity is destroyed, the surface currents associated with the field are damped with the generation of Joule heat}". 
	
The surface current is induced in a perfect conductor by the Faraday electric field $E = -dA/dt$ when the external magnetic field $H$ is changed in time $dH/dt \neq 0$. According to the Newton second law $mdv/dt = qE$ the current density $j = n_{s}qv$ in a perfect conductor with a density $n_{s}$ of the mobile  carriers of a charge $q$ should change in time $dj/dt = (n_{s}q^{2}/m)E = E/\mu_{0} \lambda_{L}^{2}$ under the action of an electric field $E$, where $\lambda _{L} = (m/\mu _{0}q^{2}n_{s})^{0.5}$  is the quantity generally referred to as the London penetration depth. According to the Maxwell equations $rot E = -dB/dt$, $rot H = j$ and $B =  \mu _{0}H$ the expression 
$$\lambda_{L}^{2}\bigtriangledown ^{2}\frac{dH}{dt} = \frac{dH}{dt}  \eqno{(1)}$$
should be valid. According to (1) a change in the magnetic field $H$ over time can penetrate in the perfect conductor only up to the penetration depth  $\lambda _{L}$. The magnetic field inside a long cylinder with a macroscopic radius $R \gg \lambda_{L}$ equals 
$$h = H_{2}\exp \frac{R-r}{\lambda_{L}}  \eqno{(2)}$$
after increasing the external magnetic field from $H = 0$ to $H = H_{2}$ at $T_{2} < T_{c}$, Fig.1. The density of the surface screening current equals
$$j = j_{0}\exp \frac{r - R}{\lambda_{L}}  \eqno{(3)}$$ 
where $j_{0} = H_{2}/\lambda_{L}$ is the current density at $r = R$ 

The screening current (3) has the kinetic energy, the density of which $\varepsilon = n_{s}mv^{2}/2 = \mu _{0}\lambda _{L}^{2}j^{2}/2$ corresponds to the energy
$$E_{k} = \mu _{0}H_{2}^{2}\frac{\lambda_{L}}{R} \eqno{(4)}$$ 
per a unit volume of a macroscopic cylinder with $R \gg \lambda_{L}$. The energy (4) is generated by the additional work performed by the power source creating the external magnetic field $H_{2}$. This energy and the screening current (3) remain in the magnetic field $H_{2}$ constant in time until the resistivity becomes non-zero $\rho > 0$. It is well known that the electric current (3) must decay within a short time and its energy must dissipate in Joule heat after the transition in the normal state with a non-zero resistivity $\rho > 0$ in the point 2' on Fig.1. Therefore, it is quite natural that physicists were sure before 1933 that "{\it when the superconductivity is destroyed, the surface current associated with the field are damped with the generation of Joule heat}" \cite{Shoenberg1952}. 

\begin{figure}
\includegraphics{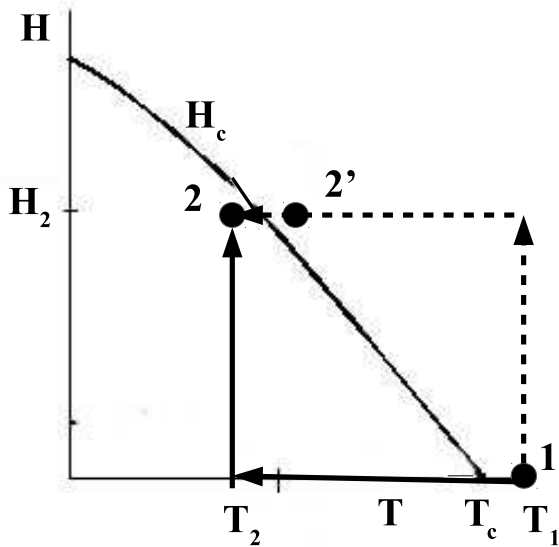}
\caption{\label{fig:epsart} A bulk superconductor or perfect conductor can go from the normal state at point 1 to superconducting or perfect conductivity state at point 2 in two ways: 1) First the temperature is lowered from $T_{1} > T_{c}$ to $T_{2} < T_{c}$ and thereafter the magnetic field is increases from $H = 0$ to $H = H_{2}$ at the temperature $T_{2} < T_{c}$; 2) First the magnetic field is increase from $H = 0$ to $H = H_{2}$ at $T_{1} > T_{c}$ and thereafter the temperature is lower from $T_{1} > T_{c}$ to $T_{2} < T_{c}(H_{2})$. The curve $H_{c}$ is the temperature dependence of the critical magnetic field.}
\end{figure} 

\section{The Meissner effect contradicts to the second law of thermodynamics if Joule heat is developed}
But if the transition from the superconducting state to the normal state is an irreversible thermodynamic process during which the kinetic energy (4) of the screening current (3) dissipates into Joule heat then the Meissner effect is experimental evidence of a process reverse to the irreversible thermodynamic process during which Joule heat is converted back into the kinetic energy (4) of the screening current (3). Such reverse process cannot be possible according to the second law of thermodynamics. Therefore physicists had to postulate after 1933 that the surface screening current (3) associated with the field are damped without the generation of Joule heat. For example, W. H. Keesom believed that "{\it it is essential that the persistent currents have been annihilated before the material gets resistance, so that no Joule-heat is developed}" \cite{Keesom1934}.  

Therefore the superconducting transition is considered by the entire superconducting community and in all theories of superconductivity \cite{Gorter1934,FHLondon1935,GL1950,BCS1957} proposed after 1933 as a phase transition, i.e. a reversible equilibrium thermodynamic process during which no Joule heating is possible. Thus, the contradiction to which Jorge Hirsch draws reader's attention \cite{Hirsch2020Physica,Hirsch2020EPL,Hirsch2020ModPhys} became possible due to the reluctance of physicists in the thirties of the last century to contradict the second law of thermodynamics. The phase transition does not contradict to the second law of thermodynamics since the free energy of the normal and superconducting states should be equal at the point of the transition $f_{n} = f_{s}$ \cite{Shoenberg1952,Gorter1934} according to the theory of the phase transition. 

It is assumed in the conventional theory of superconductivity \cite{GL1950,BCS1957} that a second-order phase transition occurs at the critical temperature $T_{c}$ since $f_{n0} < f_{s0}$ at $T > T_{c}$ and $f_{n0} > f_{s0}$ at $T < T_{c}$. The critical temperature of a bulk superconductor decreases in a magnetic field $H$, $T_{c}(H) < T_{c}(H=0)$, Fig.1, according to results of measurements \cite{Shoenberg1952,Huebener,Tink75}. Therefore the energy of the magnetic field $E_{m} = \mu _{0}H_{2}^{2}/2$ must be greater in the superconducting state, in order the equality of free energies could be possible $f_{nH} > f_{sH}$ at all values of the critical temperature $T_{c}(H)$ or critical magnetic field $H_{c}(T)$. Therefore the equality        
$$f_{n0} - f_{s0} = \frac{\mu _{0}H_{c}^{2}}{2}  \eqno{(5)}$$
in the point of the critical magnetic field $H_{c}$ was postulated \cite{Shoenberg1952,Huebener,Tink75}. 
The transition in the magnetic field $H_{c}$ of a bulk superconductor is considered after 1933 as the first-order phase transition at which the entropy jump
$$S_{n} - S_{s} = -\mu _{0}H_{c}\frac{dH_{c}}{dT}  \eqno{(6)}$$
takes place in contrast to the second-order phase transition at $H = 0$ \cite{Shoenberg1952}. 

The radical change of opinion about superconducting transition after 1933 fits with the second law of thermodynamics, but contradicts logic: the process of energy dissipation of an electric current in the normal state cannot depend on how this current appeared in the superconducting state. The bulk superconductor or the perfect conductor can reach superconducting or perfect conductivity state in the point 2 from the normal state in the point 1 on Fig.1 by two ways. The surface screening current (3) appears in the bulk superconductor due to two fundamentally different reasons on the first way and the second way. 

The current is induced both in the superconductor and in the perfect conductor by electric field $E = - dA/dt = -(2\pi r)^{-1}d\Phi /dt$ in accordance with Faraday's law on the first way when initially the temperature is lowered from $T_{1} > T_{c}$ to $T_{2} < T_{c}$ in zero magnetic field $H = 0$, and then the external magnetic field increases from $H = 0$ to $H = H_{2}$, Fig.1. Whereas on the second way this current emerges only in the superconductor and its emergence contradicts to Faraday's law and other laws of physics including the law of conservation. Therefore smart students  in the Hirsch article \cite{Hirsch2020APS} were thinking about the Meissner effect: "{\it I can't possibly see how momentum conservation is accounted for and Faraday's law is not violated}". 

Most experts in superconductivity, in contrast to the smart students and Jorge Hirsch, ignore this Meissner effect puzzle \cite{Hirsch10Meis}. Most part of the experts ignore also the question: "How can the surface current associated with the field $H_{2}$ be damped without the generation of Joule heat when the superconductivity is destroyed after the transition into the normal state?" Physicists were sure before 1933 that "{\it when the superconductivity is destroyed, the surface current associated with the field are damped with the generation of Joule heat}" \cite{Shoenberg1952} since the surface current cannot emerge according to the laws of physics in the perfect conductor at $H_{2} = H_{c}(T)$ when the temperature decreases from the point 2' to the point 2 on Fig.1. Therefore the perfect conductivity cannot contradicts to the second law of thermodynamics. Only the emergence of the surface current (3) in a magnetic field constant in time, contrary to the known laws of physics \cite{Hirsch10Meis} but discovered by  W. Meissner and R. Ochsenfeld \cite{Meissner1933}, contradicts the second law of thermodynamics, if this current is damped with the generation of Joule heat in the normal state. 

Therefore physicists postulated only after 1933 that the superconducting transition is the phase transition during which no Joule heating should be. But they did not explain how the surface current (3) can be damped without the generation of Joule heat. The conventional theory of superconductivity also cannot explain this puzzle postulated after 1933 in order to avoid the obvious contradiction with the second law of thermodynamics. Therefore Hirsch states that this theory predicts Joule heating \cite{Hirsch2020Physica,Hirsch2020EPL,Hirsch2020ModPhys}. 

Hirsch does not claim that his theory of superconductivity can explain how an electric current can be damped in the normal state without the generation of Joule heat. According to his theory "{\it Joule heating occurs only in the normal region of the material, hence no Joule heating occurs in the superconducting region}" \cite{Hirsch2020Physica}. Thus, no theory of superconductivity can explain how the surface current can be damped in the normal state without the generation of Joule heat. Hirsch does not doubt that Joule heating occurs in the normal region since "{\it Joule heating is a non-equilibrium dissipative process that occurs in a normal metal when an electric current flows, in an amount proportional to the metal's resistance}" \cite{Hirsch2020Physica}. The postulate about the reversibility of the superconducting transition in a magnetic field, made after 1933, contradicts obviously to this well known physics.             

\section{The conventional theories of superconductivity}
 \label{}
The Hirsch statement that the conventional theory of superconductivity \cite{GL1950,BCS1957} predicts the generation of Joule heat \cite{Hirsch2020Physica,Hirsch2020EPL,Hirsch2020ModPhys} has fundamental importance because this theory successfully describes numerous macroscopic quantum phenomena observed in different superconducting structures \cite{Huebener,Tink75}. Therefore, if this statement is correct, then not only the conventional theory of superconductivity \cite{GL1950,BCS1957}, but also some superconducting phenomena may contradict the second law of thermodynamics. In this regard, it is important to consider what and how the Ginsburg-Landau (GL) theory \cite{GL1950} describes, and what this theory cannot describe.

The GL wave function $\Psi _{GL} =|\Psi _{GL}|\exp{i\varphi }$ and the expression for the density of superconducting current
$$j = \frac{q}{m}n_{s}(\hbar \nabla \varphi - qA) \eqno{(7)}$$
of the superconducting mobile carriers of a charge $q$ with a mass $m$ and a density $n_{s}$ allow to describe the effects of the quantization observed in superconductors and the Meissner effect as a special case of the magnetic flux quantization. $A$ is the magnetic vector potential. Since the complex wave function $\Psi _{GL} =|\Psi _{GL}|\exp{i\varphi }$ must be single-valued at any point in the superconductor its phase $\varphi $ must change by integral multiples $2\pi$ following a complete turn along a path $l$ of integration $\oint_{l}dl \nabla \varphi = n2\pi $ \cite{Huebener}. Therefore the relation  
$$\mu _{0}\oint_{l}dl \lambda _{L}^{2} j  + \Phi = n\Phi_{0}  \eqno{(8)}$$  
between the current density $j$ along any closed path $l$ and a magnetic flux $\oint_{l}dl A = \Phi $ inside this path must be valid, according to the GL expression (7), where $\Phi _{0} = 2\pi \hbar /q$ is the quantity called flux quantum. 

The quantization of the magnetic flux $\Phi = n\Phi_{0}$ is observed \cite{fluxquan1961} in superconducting cylinders with thick walls $w \gg \lambda _{L}$, which have a closed path $l$ along which the superconducting current density is zero $\oint_{l}dl j = 0$. The quantum number $n$ can be non-zero in (8) if only a singularity of the wave function $\Psi _{GL} =|\Psi _{GL}|\exp{i\varphi }$ is inside the closed path $l$. A hole in the superconducting cylinder \cite{fluxquan1961} and the Abrikosov vortex \cite{Abrikosov} are such singularities. The integral $\oint_{l}dl \nabla \varphi = n2\pi = 0$ and the quantum number $n$ must equal zero when the closed path $l$ can be reduced to a point in the region inside $l$ without singularity. 

Therefore the Meissner effect may be considered as a special case of the flux quantization $\Phi = n\Phi_{0} $ when the quantum number $n = 0$ \cite{FPP2008}. The quantization of the superconducting current $I_{p} = sj = sqn_{s}v$ 
$$I_{p} = \frac{s}{\mu _{0}\lambda _{L}^{2}2\pi r}(n\Phi_{0}- \Phi )  = \frac{\Phi_{0}}{L_{k}}(n- \frac{\Phi }{\Phi_{0}}) \eqno{(9)}$$
is observed in the case of a weak screening in a cylinder with walls $w \ll \lambda _{L}$ \cite{LP1962} or a ring with the cross-section $s \ll \lambda _{L}^{2}$ \cite{Physica2019,PLA2012Ex,Letter2007,JETP2007,Letter2003}. $L_{k} = (\lambda _{L}^{2}/s)\mu_{0}2\pi r = m2\pi r/sq^{2}n_{s} $ is the kinetic inductance of the ring with the radius $r$, the gross-section $s$ and the density $n_{s}$ of superconducting particles with a charge $q$.

\section{The conventional theories do not describe dynamic processes in superconductors}
The GL theory \cite{GL1950} describes successfully both the Meissner effect and the persistent current (9) observed in a superconducting ring with a weak screening $s \ll \lambda _{L}^{2}$ \cite{Physica2019,PLA2012Ex,Letter2007,JETP2007,Letter2003} as a consequence of the quantization of the canonical momentum $p = mv + qA$   
$$\oint_{l}dl p = \oint_{l}dl (mv + qA) = \oint_{l}dl \hbar \nabla \varphi =  n2\pi \hbar \eqno{(10)}$$ 
But neither the GL theory \cite{GL1950} nor the BCS theory \cite{BCS1957} say anything about the 'force' which could change the momentum of mobile charge carriers during the transition of the bulk superconductor or the ring between the normal and superconducting states. 

The Hirsch statement that the conventional theory of superconductivity predicts Joule heating \cite{Hirsch2020Physica,Hirsch2020EPL,Hirsch2020ModPhys} implies that according to this theory \cite{GL1950,BCS1957} the surface screening current (3) and the persistent current (9) are damped in the normal state under influence of the dissipation force. No direct prediction of Joule heating is in the GL theory \cite{GL1950} and the BCS theory \cite{BCS1957}. The presence of such prediction in the theories created within the framework of equilibrium thermodynamics would mean that their authors did not know that Joule heating is an irreversible thermodynamic process. Nevertheless the dissipation force is implied since neither the GL theory \cite{GL1950} nor the BCS theory \cite{BCS1957} proposed an alternative force. 

The example of the persistent current (9) in a superconducting ring with a weak screening $s \ll \lambda _{L}^{2}$ has an advantage in comparison with the screening current (3) in a bulk superconductor for the analysis of GL theory \cite{GL1950} because of two reasons: 1) all superconducting pairs have the same velocity in the ring with $s \ll \lambda _{L}^{2}$ and 2) the quantum number $n$ describing the canonical momentum (10) of the pairs can have different values in the ring, whereas $n = 0$ in the Meissner state. The magnetic inductance of such ring $L_{f} \approx \mu_{0}2\pi r \approx (s/\lambda _{L}^{2})L_{k} \ll L_{k}$ is smaller then the the kinetic inductance $L_{k} = (\lambda _{L}^{2}/s)\mu_{0}2\pi r $ and the magnetic flux induced by the persistent current is small $\Delta \Phi _{I} = L_{f}I_{p} \ll \Phi = BS$ \cite{QuanStud2016}.   

The persistent current (9), as well as the screening current (3), can appear due to two fundamentally different reasons when the maximum magnetic flux inside the ring is enough low $\Phi = BS < 0.5\Phi_{0}$. According to the GL theory the density of the superconducting particles increases from $n_{s} = 0$ at $T_{1} > T_{c}$ to $n_{s} = n_{s}(0)(1-T_{2}/T_{c})$ at $T_{2} < T_{c}$ when the temperature is lowered on the first way, Fig.1, in the zero magnetic field $H = 0$. The velocity increases from $v = 0$ to $v = -q\Phi /m 2\pi r = -(\hbar /mr)\Phi /\Phi_{0}$ under influence of the Faraday electric field $E = -dA/dt = -(2\pi r)^{-1}d\Phi /dt$ in accordance with the Newton second law $mdv/dt = qE$ when the external magnetic field increase on the first way, Fig.1, from $H = 0$ to $H = H_{2} = \Phi /\mu_{0}S$ at the temperature $T_{2} < T_{c}$. The persistent current (9) increases from $I_{p} = 0$ to $I_{p} = -\Phi /L_{k}$ in accordance with the classical electrodynamics law $L_{k}dI_{p}/dt = -d\Phi /dt$. 

The GL theory describes this dynamical process in which the velocity is changed in accordance with the laws of classical physics. But neither the GL theory, nor any other theory of superconductivity can describe the dynamics of the emergence of the persistent current $I_{p} = -\Phi /L_{k}$ on the second way, Fig.1, when the temperature is lowered from $T_{1} > T_{c}$ to $T_{2} < T_{c}$ in the non-zero magnetic field $H_{2}$. The GL theory predicts that the persistent current (9) $I_{p} = sj = sqn_{s}v$ should appear at $n_{s} > 0$ because the velocity 
$$\oint_{l}dl v  =  \frac{2\pi \hbar }{m}(n - \frac{\Phi}{\Phi_{0}}) \eqno{(11)}$$     
cannot be zero at $\Phi \neq n\Phi_{0}$ according to the quantization of the canonical momentum (10). 

According to the GL theory all $N_{s} = n_{s}V$ superconducting particles have the same quantum number $n$ in the superconducting ring with the volume $V = s2\pi r$ and $n = 0$ in the bulk superconductor. Therefore the persistent current $I_{p} = sqn_{s}v = qN_{s}v/2\pi r$ and the density of the surface screening current (3) equal $j \approx H_{2}(R-r)/\lambda_{L}^{2} = H_{2}(R-r)\mu _{0}q^{2}n_{s}/m \propto n_{s}$ at $R-r \ll \lambda _{L}$ are proportional to the density of the superconducting particles equal $n_{s} = n_{s}(0)(1-T/T_{c})$ at $T < T_{c}$. The GL theory predicts that the velocity (11) and the persistent current (9) depend not only on the value of the magnetic field $H_{2} = \Phi /\mu_{0}S$ but also on the quantum number $n$. Therefore the persistent current (9) may take different discrete values $I_{p} = I_{p,A}2(n - \Phi /\Phi_{0})$ at the same value $H_{2} = \Phi /\mu_{0}S$ after the transition of the ring in the superconducting state with different quantum number $n$: $I_{p}(n+1) - I_{p}(n) = I_{p,A}2 = \Phi _{0}/L_{k} = sn_{s}q\pi \hbar/rm$.  

The spectrum of the permitted values of the kinetic energy 
$$E_{k} = \frac{L_{k}I_{p}^{2}}{2} =  I_{p,A}\Phi_{0}(n - \frac{\Phi }{\Phi_{0}}) ^{2} \eqno{(12)}$$ 
is strongly discrete according to the GL theory. The difference $|E_{k}(n+1) - E_{k}(n)| \approx 0.6 I_{p,A}\Phi_{0}$ at $\Phi \approx (n+0.2)\Phi_{0}$ of a superconducting ring with a typical value $I_{p,A} \approx 10 \ \mu A$ of the persistent current \cite{JETP2007} corresponds the temperature $0.6 I_{p,A}\Phi_{0}/k_{B} \approx 1000 \ K$ \cite{NanoLet2017}. Therefore the GL theory predicts that the quantum number $n$ with the minimum kinetic energy (12) has the predominate probability $P(n) \propto \exp -E_{k}(n)/k_{B}T$. This number changes with the magnetic flux $\Phi = BS = \mu_{0}HS$ inside the ring: $n = 0$ at $-0.5\Phi_{0} < \Phi < 0.5\Phi_{0}$; $n = 1$ at $0.5\Phi_{0} < \Phi < 1.5\Phi_{0}$; $n = 2$ at $1.5\Phi_{0} < \Phi < 2.5\Phi_{0}$ etc. Thus, the GL theory predicts that the value and the direction of the persistent current should change periodically in magnetic field with the period $H_{0} = \Phi _{0}/\mu_{0}S$ corresponding to the flux quantum $\Phi _{0}$ inside the ring.
        
All measurements, for example, of the critical current \cite{JETP2007,JETP07J,PLA2020} corroborate this prediction. The measured values of the critical current \cite{JETP2007,JETP07J,PLA2020} correspond to the predominate probability of the permitted state $n$ with the minimal kinetic energy (12) at a give value of the magnetic flux $\Phi = BS$ in the most cases. The observation of two states $n$ and $n+1$ at the same $\Phi = BS$ value in rare cases, see Fig.1 in \cite{PLA2020}, only emphasizes the strong discreteness of the permitted state spectrum of superconducting rings. The critical current \cite{JETP2007,JETP07J,PLA2020} and other measured parameters \cite{Physica2019,PLA2012Ex,Letter2007,JETP2007,Letter2003} connected with the persistent current (9) change periodically in magnetic field $B$ with the period $H_{0} = \Phi _{0}/\mu_{0}S$ corresponding to the flux quantum $\Phi _{0} = 2\pi \hbar /q$ inside the ring with $S = \pi r^{2}$. The value of the period $\Phi _{0} \approx 20.7 \ Oe \ \mu m^{2}$ gives experimental evidence that the persistent current in superconductors is the current of pairs of electrons with the charge $q = 2e$.        

Measurements corroborate also the temperature dependence of the amplitude $I_{p,A} = sn_{s}q\pi \hbar/2rm \propto n_{s} = n_{s}(0)(1-T/T_{c})$ of the oscillation of the persistent current and of the density of the surface screening current (3) $j  \propto n_{s}= n_{s}(0)(1-T/T_{c})$ predicted by the GL theory. The GL theory describes also other numerous quantum phenomena observed in superconductors. But neither the GL theory, nor any other theory of superconductivity can describe the dynamics of the change of the velocity (11) because of the change of the quantum number $n$ and the transition between normal and superconducting states. The temperature dependence $I_{p} \propto n_{s} = n_{s}(0)(1-T/T_{c})$ implies that the velocity $v$ of the mobile charge carriers should jump between $v = 0$ and the quantum value (11) $v  =  (\hbar /mr)(n - \Phi /\Phi_{0})$ at their transition between superconducting and normal states. The angular momentum of the macroscopic number $N_{s} = n_{s}s2\pi r$ of the mobile charge carriers should change on the macroscopic value $N_{s}\hbar (n - \Phi /\Phi_{0})$ at the transition the ring with a macroscopic radius $r$ and a macroscopic cross-section $s$ from normal to superconducting state. Neither the GL theory, nor any other theory of superconductivity say anything about the puzzle of the jump of the angular momentum on the macroscopic value $N_{s} r\Delta p \approx  (10^{5} \div 10^{10})\hbar $ which this theory predicts and which is observed \cite{NanoLet2017}. 

\section{What is the 'force' propelling the mobile charge carriers?}
Jorge Hirsch expressed astonishment that this puzzle is ignored: "{\it Strangely, the question of what is the 'force' propelling the mobile charge carriers and the ions in the superconductor to move in direction opposite to the electromagnetic force in the Meissner effect was essentially never raised nor answered to my knowledge, except for the following instances: \cite{London1935} (H. London states: "The generation of current in the part which becomes supraconductive takes place without any assistance of an electric field and is only due to forces which come from the decrease of the free energy caused by the phase transformation," but does not discuss the nature of these forces), \cite{PRB2001} (A.V. Nikulov introduces a 'quantum force' to explain the Little-Parks effect in superconductors and proposes that it also explains the Meissner effect)}" \cite{Hirsch10Meis}. 

It should be noted here that H. London in 1935 was sure that the transition from the superconducting to the normal state is an equilibrium reversible process that can be described by the free energy, although before 1933 physicists were sure that this transition is an irreversible thermodynamic process with the generation of Joule heat \cite{Shoenberg1952}. I should note that the "quantum force" introduced in \cite{PRB2001} describes rather than explains the Little-Parks effect and it cannot claim to explain the Meissner effect. 

The "quantum force" describes only the change of the velocity from $v = 0$ to the quantum value (11) $v  =  (\hbar /mr)(n - \Phi /\Phi_{0})$ predicted by the GL theory and observed experimentally. The angular momentum of superconducting pair equals $rp = rmv + rqA = rmv + q\Phi /2\pi = rmv + \hbar \Phi /\Phi _{0} = n\hbar $ in the ring with the radius $r$ according to quantum mechanics, whereas the average velocity of electron pair equals zero $v = 0$ and $rp =  \hbar \Phi /\Phi _{0}$. Thus, quantum mechanics states that the angular momentum of each electron pair changes on $  \hbar (n - \Phi /\Phi _{0})$ at each transition of the ring into the superconducting state. The total change during a time unity of the momentum 
$$F_{q} = \hbar (\overline{n} - \frac{\Phi }{\Phi_{0}})f_{sw}/r \eqno{(13)}$$ 
when the ring is switched between the superconducting and the normal state with a frequency $f_{sw} = N _{sw}/\Theta$ was called 'quantum force' in \cite{PRB2001}. $\overline{n} = \int_{\Theta} dt n/\Theta = \sum _{i=1}^{i=N _{sw}}n_{i}/N _{sw}$ is the average value of the quantum number after $N _{sw} \gg 1$ comeback of the ring into the superconducting state.         

The quantum force was introduced in \cite{PRB2001} in order to describe the Little-Parks effect and other experimental evidences of the persistent current $I_{p} \neq 0$ which does not decay at a non-zero resistance $R > 0$. W.A. Little and R.D. Parks called their article "Observation of Quantum Periodicity in the Transition Temperature of a Superconducting Cylinder" \cite{LP1962}. Therefore the Little-Parks effect is considered as a depression of the transition temperature $T_{c}$ because of a non-zero velocity (11) or the persistent current (9) at $\Phi \neq n\Phi_{0}$ \cite{Tink75}, although the magnetic dependence of the critical temperature $T_{c}(\Phi)$ was measured only in a few works \cite{Moshchalkov1992}. 

The quantum periodicity of the resistance $R(\Phi)$ rather than of the transition temperature was observed in \cite{LP1962} and most other publications, for example \cite{Letter2007}. Thus, the Little-Parks effect testifies that the persistent current $I_{p} \neq 0$ can be observed in a cylinder \cite{LP1962} and a ring \cite{Letter2007} with a non-zero resistance $R > 0$. This paradoxical phenomena is observed in a narrow temperature region corresponding the resistive transition from the normal to superconducting state where the resistance of the ring greater than zero $R > 0$, but less than the resistance in the normal state $R < R_{n}$. The resistance $R = R_{n}$ but $I_{p} = 0$ above this region, whereas below this region $I_{p} \neq 0$ but $R = 0$. 

The deducing of the quantum force \cite{PRB2001} does not go beyond the GL theory and quantum mechanics. Therefore, the publication \cite{PRB2001} cannot assert anything that these theories do not assert. It is assumed both in the GL theory and \cite{PRB2001} that the angular momentum of each pair of electrons changes from $rp = n\hbar $ to $rp =  \hbar \Phi /\Phi _{0}$ in the normal state under influence of the dissipation force. The dissipation force acts between the mobile charge carriers and the ions of the normal metal. Therefore it is assumed in \cite{PRB2001} that the quantum force acts also between the mobile charge carriers and the ions. Otherwise the superconducting cylinder \cite{LP1962} or the ring \cite{Letter2007} would start rotating in the Little-Parks experiment. Neither the GL theory nor quantum mechanics say anything about the force that changes the angular momentum of the mobile charge carriers and the ions when the persistent current (9) emerges in the superconducting state of the ring.      

\section{Experimental evidence of the generation of Joule heat}
W.H. Keesom believed it is necessary "{\it that the persistent currents have been annihilated before the material gets resistance, so that no Joule-heat is developed}" \cite{Keesom1934}. But the observations of the Little-Parks effect \cite{LP1962,Letter2007} and the direct observation \cite{PC2007} of the persistent current near superconducting transition $T \approx T_{c}$ where the ring resistance $0 < R < R_{n}$ give experimental evidence that the persistent current is not annihilated even after the material gets resistance $R > 0$. Measurements give evidence that the persistent current can even induce the dc voltage $V _{dc}(\Phi ) \propto \overline{I_{p}}(\Phi )$, see Fig.4 in \cite{Physica2019}. Thus, the radical change of the opinion about superconducting transition after 1933 contradicts not only logic but also the experimental results. The observations \cite{LP1962,Letter2007,PC2007} of the persistent current $I_{p} \neq 0$ which is not damped at a non-zero resistance $0 < R < R_{n}$ in the absence of the Faraday electric field $E = -dA/dt = -(2\pi r)^{-1}d\Phi /dt = 0$ is obvious paradox. But this paradox can be described in the framework of the conventional theory of superconductivity \cite{GL1950}.    

The density of electric current $j = qn_{e}\overline{v}$ is non-zero when the average velocity of the mobile charge carriers is not zero $\overline{v} \neq 0$. The dissipation force $F_{dis} = -\eta \overline{v}$ acts in any normal metal because of electron scattering. The electron scattering results to a non-zero resistivity $\rho \propto 1/l_{fp}$ which is inversely proportional to the mean free path $l_{fp}$ between the scattering. Therefore the conventional circular electric current $I \neq 0$ can be non-zero in a ring with a non-zero resistance $R = \rho 2\pi r/s > 0$ if only the dissipation force is balanced $F_{dis} + qE = 0$ by the force $qE$ of the electric field $E$. The equality $RI = \rho 2\pi r j = 2\pi r E = - d\Phi /dt$ for the electric current constant in time $dI/dt = 0$ is deduced from the force balance $F_{dis} + qE = 0$. The force balance is $m\overline{v}/dt = F_{dis} = -\eta \overline{v}$ when the Faraday electrical field $E 2\pi r = -d\Phi /dt = 0$ and the conventional current must quickly decay in this case during a relaxation time $\tau_{RL} = L/R$. 

The current does not decay in the state of the perfect conductivity or superconductivity because of the absence of the scattering of the superconducting particles the mean free path of which is infinite $l_{fp} = \infty $. The decrease of the persistent current $I_{p} \propto n_{s} = n_{s}(0)(1-T/T_{c}) \rightarrow 0$ at $T \rightarrow T_{c}$ implies that the mean free path $l_{fp}$ of particles becomes finite after their transition into the normal state. The average velocity $\overline{v}$ of non-superconducting electrons decreases down to zero under the influence of the dissipation force $F_{dis} = -\eta \overline{v}$. Joule heating occurs during this dynamical process. The conventional theory of superconductivity \cite{GL1950} does not consider directly the dissipation force. But Joule heating is implicitly assumed since according to this theory the persistent current $I_{p} = sqn_{s}v =  \propto n_{s} = n_{s}(0)(1-T/T_{c})$ is proportional to the density of superconducting pairs which decreases with the temperature increase.  

The thermal fluctuations \cite{Tink75} switch the ring or its segments between the normal state with $n_{s} = 0$ and $R = R_{n}$ and the superconducting state with $n_{s} > 0$ and $R = 0$. Therefore the persistent current $\overline{I_{p}} \neq 0$ is observed at a non-zero resistance average in time $\overline{R} > 0$ in a narrow fluctuation region near the temperature of superconducting transition $T \approx T_{c}$ \cite{LP1962,Letter2007,PC2007}. The persistent current does not decay at $d\Phi /dt = 0$ and $\overline{R} > 0$ because the quantum force (13) balances the dissipation force $F_{q} + \overline{F_{dis}} = 0$. The angular momentum changes from $rp = n\hbar $ to $rp = \hbar \Phi /\Phi _{0}$ in the normal state with $R > 0$ under influence of the dissipation force $F_{dis} = -\eta \overline{v}$. The dissipation force acts at each transition of the ring in the normal state since the angular momentum returns to the quantum value $rp = n\hbar $ at each transition into the superconducting state. Therefore the dissipation force average over a time $\Theta \gg 1/f _{sw}$ should be equal $\overline{F_{dis}} = \int_{\Theta} dt F_{dis}/\Theta = \hbar (\Phi /\Phi_{0}- \overline{n})f_{sw}/r $ to the quantum force with the inverse sign $\overline{F_{dis}} = - F_{q}$. The persistent current $\overline{I_{p}} \neq 0$ does not decay at $\overline{R} > 0$ \cite{Letter2007,LP1962,PC2007} despite the energy dissipation with the power $\overline{RI^{2}} \neq 0$ since the quantum force $F_{q}$ replaces the force $qE$ of the Faraday electrical field $E 2\pi r = -d\Phi /dt $. 

The persistent current $\overline{I_{p}} \neq 0$ is observed at $\overline{R} > 0$ not only in the fluctuation region near superconducting transition $T \approx T_{c}$ \cite{LP1962,Letter2007,PC2007} but also in normal metal rings \cite{Science2009PC,PRL2009PC}. The authors \cite{Science2009PC} claim that the persistent current which they observed in normal metal rings can be dissipationless. The author \cite{Birge2009} agrees with them, but recognizes: "{\it The idea that a normal, nonsuperconducting metal ring can sustain a persistent current - one that flows forever without dissipating energy - seems preposterous. Metal wires have an electrical resistance, and currents passing through resistors dissipate energy}". This idea not only is preposterous but also contradicts to elementary mathematics. The authors \cite{Science2009PC} observe $I_{p} \neq 0$ and measure $R > 0$, but they claim that $RI_{p}^{2} = 0$. No one can claim that the current $I$ induced by the Faraday voltage $RI = -d\Phi /dt$ in a ring with a resistance $R > 0$ is dissipationless. No one claimed that the Nyquist \cite{Nyquist} (or Johnson \cite{Johnson}) noise current $\overline{I_{Nyq}^{2}}  = k_{B}T\Delta \omega /R$ observed in a ring with $R > 0$ under thermodynamic equilibrium in a frequency band $\Delta \omega $ is dissipationless.

\section{The Nyquist current and the persistent current}
Both the Nyquist current and the persistent current are observed in the rings with non-zero resistance \cite{Letter2007,PC2007,Science2009PC,PRL2009PC} at thermodynamic equilibrium due to thermal fluctuations. The persistent current both of superconducting pairs \cite{Letter2007,PC2007} and electrons \cite{Science2009PC,PRL2009PC} differs from the Nyquist current $\overline{I_{Nyq}^{2}}  = k_{B}T\Delta \omega /R$ because of a non-zero value at the zero frequency $\omega = 0$. The directed current $\overline{I_{p}}  = \int_{\Theta} dt I_{p}/\Theta \neq 0$  is observed \cite{Letter2007,PC2007,Science2009PC,PRL2009PC} due to the discreteness of the permitted states (11) when the quantization (10) takes place. This current $\overline{I_{p}} = \overline{I_{p,A}2(n - \Phi /\Phi_{0})} = \overline{I_{p,A}}2\sum _{n}(n - \Phi /\Phi_{0})P(n)$ can be non-zero when one of the permitted states $n$ has the predominate probability $P(n) \propto \exp -E_{k}(n)/k_{B}T$, i.e. when the energy difference $|E_{k}(n+1) - E_{k}(n)|$ between the permitted states $n$ and $n+1$ is not much less than $k_{B}T$.  

The period $\Phi _{0} = 2\pi \hbar /e \approx 41.4 \ Oe \ \mu m^{2}$ of the quantum oscillation observed at measurements of normal metal rings \cite{Science2009PC,PRL2009PC} give experimental evidence that the persistent current in this case is the current of electrons with the charge $q = e$. The superconducting ring has a strong discreteness of the spectrum of the permitted states (12), for example $|E_{k}(n+1) - E_{k}(n)|/k_{B} \approx 1000 \ K$ at $r \approx 2 \ \mu m = 2 \ 10^{-6} \ m$ and $I_{p,A} \approx 10 \ \mu A$ \cite{JETP2007}, since the macroscopic number $N_{s} = n_{s}s2\pi r$ of pairs have the same quantum number $n$ \cite{PLA2020}. The discreteness of electrons $E_{n+1} - E_{n} = (2n+1)\hbar ^{2}/2mr^{2}$ corresponds to the much lower temperature $\hbar ^{2}/2mr^{2}k_{B} \approx 0.001 \ K$ in a normal metal ring with the radius $r \approx 500 \ nm = 5 \ 10^{-7} \ m$ at $n = 0$ \cite{PLA2020}. 

Electrons, being fermions, occupy the levels from $n = -n_{F}$ to $n = +n_{F}$ with the opposite direction of the velocity and therefore the persistent current of electrons $I_{p,1} \approx (e/2\pi r)v_{F} \approx (e\hbar/m2\pi r^{2})n_{F} $ is created by one electron on the Fermi level $n _{F}$ per one - dimensional channel \cite{PRB1988tPCNM}. The persistent current of electrons is observed in normal metal nano-rings with a radius $r > 300 \ nm$ at the temperature $T \approx 1 \ K$ \cite{Science2009PC,PRL2009PC} because the quantum number at the Fermi level is very great $n_{F} \gg 1$ and therefore $(2n+1)\hbar ^{2}/2mr^{2}k _{B} \approx 1 \ K$. 

The resistivity of normal metal is not zero $\rho > 0$ and therefore the energy of electric current with a density $j$ is dissipated into Joule heat with a power density $\rho j^{2}$ since the mean free path $l_{fp}$ between the scattering is finite. The persistent current, in contrast to the conventional electric current, does not decay without the electric force $E = 0$ in normal metal rings \cite{Science2009PC,PRL2009PC} since electrons return after scattering from time to time to the quantum state (10) and their average velocity becomes non-zero (11), like the velocity of superconducting pairs, when the magnetic flux inside the ring is not divisible by the flux quantum $\Phi \neq n\Phi_{0}$. The change of the angular momentum of electrons because of the quantization (10) during a time $\Theta $ may be also describe by the quantum force (13).  

Thus, the persistent current does not decay at $E = 0$ in spite of the non-zero energy dissipation $\rho j^{2}$ both in superconductor and normal metal rings since the dissipation force is balanced by the quantum force $F_{dis} + F_{q} = 0$ in the both cases. The average value of the Nyquist current, as a type of the classical Brownian motion \cite{Feynman}, equals zero $\overline{I_{Nyq}}  = 0$ since the average value of a rapidly fluctuating force in the Langevin equation \cite{Langevin1908} is zero in the classical case. The quantum force is the Langevin force in the case of the persistent current observed in rings with non-zero resistance at thermodynamic equilibrium the average value of which is not zero due to the discreteness of the permitted states spectrum (13).  

The preposterous claim, made by the authors \cite{Science2009PC,Birge2009} in spite the obvious contradiction with the elementary mathematics, and the radical change of the opinion about superconducting transition after 1933 have the same reason - the belief in the second law of thermodynamics. If the dissipation power $RI_{p}^{2}$ is not zero then the observations \cite{Letter2007,LP1962,PC2007,Science2009PC,PRL2009PC} of the persistent current $\overline{I_{p}} \neq 0$ at a non-zero resistance $\overline{R} > 0$ contradict to the second law of thermodynamics. Likewise, if "{\it the surface current associated with the field are damped with the generation of Joule heat}" \cite{Shoenberg1952} then the Meissner effect contradicts to the second law of thermodynamics. Direct evidence of a violation of the second law of thermodynamics is the observation of the dc potential difference $V _{dc}(\Phi ) \propto \overline{I_{p}}(\Phi )$ on the halves of a ring with different resistance $R_{B} > R_{A}$ and the persistent current $\overline{I_{p}} \neq 0$. The dc power $V _{dc}^{2}/R$ can be easy used for an useful work in contrast to the chaotic power$V_{Nyq}^{2}/R$ of the the Nyquist noise \cite{Nyquist,Johnson,Feynman}.                          

\section{Dynamic processes in asymmetric superconducting rings}
The conventional theory of superconductivity \cite{GL1950} predicts a possibility to observe the dc voltage $V _{dc}(\Phi ) \propto \overline{I_{p}}(\Phi )$ \cite{LTP1998} and experimental results \cite{Physica2019,PLA2012Ex,Letter2007,JETP2007,Letter2003} corroborate this prediction. The quantum force $F_{q}$ replaces the Faraday force $qE_{F}$ \cite{PLA2012} in the case of the persistent current $I_{p} \neq 0$ observed in the rings with non-zero resistance $R > 0$ \cite{Letter2007,PC2007,Science2009PC,PRL2009PC}. The potential difference
$$ V = 0.5(R_{B} - R_{A})I \eqno{(14)}$$ 
is observed on the halves of the ring with different resistance $R_{B} > R_{A}$ when the conventional circular electric current $I$ is induced by the Faraday electric field $RI = (R_{B} + R_{A})I = -d\Phi /dt$. The potential difference should equal $V = 0.5R_{B}I = -0.5d\Phi /dt$ when the half $A$ is in the superconducting state $R_{A} = 0$, Fig.2, and $dI/dt = 0$. The half $A$ with $R_{A} = 0$ may be considered as the secondary winding of the electric transformer and the half $B$ with $R_{B} > 0$ as a load \cite{Physica2019} in which the energy dissipation with the power $VI = 0.5R_{B}I^{2}$ takes place.  

\begin{figure}
\includegraphics{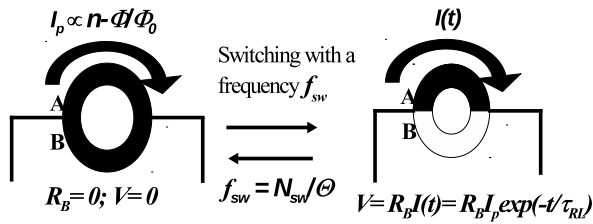}
\caption{\label{fig:epsart} The left picture: The persistent current (9) flows when the ring in superconducting state (marked in black). The right picture: the potential difference $V(t) = R_ {B}I(t)$ should be observed during the relaxation time $\tau_{RL} = L /R_ {B}$ after the transition of the half B to the normal state (indicated in white). The persistent current (9)  must reappear when the half B returns to the superconducting state, since the state with $I_{p} = 0$ is forbidden  according to the quantization condition (10) when the magnetic flux inside the ring is not divisible by the flux quantum $\Phi \neq n\Phi_ {0}$.}
\end{figure} 
 
According to the quantization (10) the persistent current (9) should appear after the transition of the half $B$ in the superconducting state at $H_{2} \neq nH_{0} = n\Phi _{0}/\mu_{0}S$ constant in time $dH_{2}/dt = 0$ because the superconducting state with zero velocity (11) is forbidden at $\Phi = H_{2}\mu_{0}S \neq n\Phi _{0}$. The kinetic inductance equals $L_{k} = ml\overline{(s n_{s})^{-1}}/q^{2}$ in the ring with the section area $s$ or the density of superconducting particles $n_{s}$ varying along the circumference $l = 2\pi r$, where $\overline{(s n_{s})^{-1}} = l^{-1}\oint _{l}dl (s n_{s})^{-1}$ \cite{Physica2019}. The circular current $I(t)$ should decay because of the dissipation force in the half $B$ and because of the potential difference 
$$V(t) = R_ {B}I(t) = R_ {B}I_{p}\exp -{\frac{t }{\tau_{RL}}}  \eqno{(15)}$$ 
in the superconducting half $A$, after the transition of the half $B$ in the normal state, Fig.2. The relaxation time $\tau_{RL} = L /R_ {B}$ is determined by the resistance $R_ {B}$ and the total inductance $L$ of the ring. We cannot doubt because of all experimental results and quantization (10) that the persistent current (9) must reappear when the half $B$ returns to the superconducting state although we cannot describe this dynamical process. The voltage (15) should be observed again after the transition of the half $B$ in the normal state. The voltage average during a time $\Theta \gg 1/f _{sw}$ should be 
$$V _{dc} =  \int _{0}^{\Theta }dt \frac{V(t)}{\Theta }= R _{B}\overline{I_{p}}f _{sw} \int _{0}^{t _{n}}dt \exp {-\frac{t}{\tau_{RL}}}  \eqno{(16)}$$ 
when the half $B$ is switched between superconducting and normal states with a frequency $f _{sw} = N _{sw}/\Theta $. Here $\overline{I_{p}}= \sum^{i=N _{sw}}_{i=1} I_{p,i}/N _{sw}$  is the average value of the persistent current after $N _{sw}$ returning of the half $B$ in the superconducting state during the time $\Theta \gg 1/f _{sw}$;  $t _{n}$ is the time during which the half $B$ is in the normal state; the integral $\int_{0}^{t_{n}} dt\exp {-t /\tau_{RL}} \approx  \tau_{RL} = L /R_{B}$ when $t_{n} \gg \tau_{RL}$ and $\int_{0}^{t_{n}} dt\exp {-t /\tau_{RL}} \approx t_{n}$ when $t_{n} \ll \tau_{RL}$. 

The average value of the persistent current $\overline{I_{p}}$ oscillates in magnetic field with the period corresponding the flux quantum $\Phi_{0}$ and changes direction at  $\Phi = n\Phi_{0}$ and $\Phi = (n+0.5)\Phi_{0}$ \cite{PC2007}. A possibility to observed the dc voltage (16)  $V _{dc}(\Phi ) \propto \overline{I_{p}}(\Phi )$, oscillating in magnetic field, when the same ring segment is switched between superconducting and normal states with an enough high frequency $f _{sw}$ was considered first in 1998 \cite{LTP1998}. Exactly such an experiment has not yet been done. The quantum oscillations $V _{dc}(\Phi ) \propto \overline{I_{p}}(\Phi )$ was observed at measurements of an asymmetric superconducting quantum interference device \cite{Physica1967} thirty years before considering the possibility of such oscillations \cite{LTP1998}. The similar oscillations at measurements of asymmetric superconducting ring were observed first in 2002 \cite{NANO2002}. The oscillations of the dc voltage $V _{dc}(\Phi ) \propto \overline{I_{p}}(\Phi )$ are observed when the asymmetric rings are switched between superconducting and normal state by non-equilibrium noise \cite{Letter2007,Physica1967,NANO2002,APL2016,Kulik75}, the external ac current \cite{Letter2003,JETP2007} or thermal fluctuations \cite{PLA2012Ex}. The observations $V _{dc}(\Phi ) \propto \overline{I_{p}}(\Phi )$ \cite{Physica2019,PLA2012Ex,Letter2007,JETP2007,Letter2003,Physica1967,NANO2002,APL2016,Kulik75} give evidence that the persistent current induces the potential difference like that the conventional circular current (14).   

The measurements \cite{Physica2019,PLA2012Ex,Letter2007,JETP2007,Letter2003,Physica1967,NANO2002,APL2016,Kulik75} of the dc voltage $V_{dc} \propto I_{p}$ indicate the observation of a dc power source $V_{dc}I_{p}$ or $V_{dc}^{2}/R_{s}$ since the persistent current flows against the dc voltage in one of the ring halves \cite{QuanStud2016} like in the case of the conventional circular electric current (14). Analogy with the electric transformer may be useful here \cite{Physica2019}. The current $I$ flows against the potential difference $V $ in the ring half with a lower resistance $R_{A}$ (14), like in the secondary winding of the electric transformer and the half with a higher resistance $R_{B}$ may be considered as the load \cite{Physica2019}. The conventional circular current $I$ flows against the potential electric field $E_{p} = -\nabla V$ but not against the total electric field $E_{t} = E_{p} + E_{F} =  -\nabla V - dA/dt$ since $I = 0$ at $R_{B} + R_{A} > 0$ and $d\Phi /dt = 0$. 

The puzzle, connected with the Meissner effect puzzle \cite{Hirsch10Meis}, is that according to the experimental results \cite{Physica2019,PLA2012Ex,Letter2007,JETP2007,Letter2003,Physica1967,NANO2002,APL2016,Kulik75} the persistent current can flow against the action of the total dc electric field $E_{t} = E_{p} =  -\nabla V_{dc}$ \cite{Physica2019,QuanStud2016} since the direct current $I_{p}$ is observed in a magnetic field constant in time $d\Phi /dt = 0$, i.e. without the Faraday electrical field $E_{F} = - dA/dt = 0$. According to this experimental puzzle the half with a lower resistance of an asymmetric ring with the persistent current may be considered as the secondary winding of the electric transformer which is a power source $V_{dc}I_{p} = V_{dc}^{2}/R_{s}$ without the primary winding \cite{Physica2019}.

\section{Conclusion}
Arthur Eddington wrote: "{\it The second law of thermodynamics holds, I think, the supreme position among the laws of Nature. If someone points out to you that your pet theory of the universe is in disagreement with Maxwell's equations - then so much the worse for Maxwell's equations. If it is found to be contradicted by observation, well, these experimenters do bungle things sometimes. But if your theory is found to be against the second law of thermodynamics I can give you no hope; there is nothing for it but collapse in deepest humiliation}" \cite{Eddington}. Physicists sacrificed logic after 1933 and the authors \cite{Science2009PC,Birge2009} made the preposterous claim contradicting to the elementary mathematics because of this supreme position of the second law of thermodynamics. 

But it was not enough to sacrifice logic and to postulate that the persistent currents have to be "{\it annihilated before the material gets resistance, so that no Joule-heat is developed}" \cite{Keesom1934}. It was necessary to explain how the surface current associated with the field can be damped without the generation of Joule heat. No one before Jorge Hirsch \cite{Hirsch2020Physica,Hirsch2020EPL,Hirsch2020ModPhys} had noticed that no theory of superconductivity could explain how an electric current could be damped in the normal state without the generation of Joule heat. 

Moreover, experts on superconductivity stopped with time to notice that the Meissner effect contradicts the second law of thermodynamics if "{\it the surface currents associated with the field are damped with the generation of Joule heat}" \cite{Shoenberg1952}. They are sure that superconducting transition is the phase transition, although the equality of the free energy (5) is doubtful because of the contradiction with the laws of electrodynamics. According to these laws the work performed by the power source to create the magnetic field $H = H_{2}$ inside the solenoid increases with the increase the magnetic permeability $\mu $ of the material inside the solenoid, since the voltage to the source must be $V > -d\Phi /dt = -SdB/dt = -S\mu \mu _{0}dH/dt$ in order the field could increases in time. 

The second law of thermodynamics holds the supreme position because of the centuries-old belief of scientists in the impossibility of a perpetuum mobile. The Carnot principle which we call "{\it as the second law of thermodynamics since Clausius's time}" \cite{Smoluchowski} was postulated on the base of this belief. Perpetuum mobile would be inevitable according to the law of energy conservation if all the laws of physics were reversible. Therefore Carnot postulated in 1824 that the conversion of the kinetic energy of any machine into heat is an irreversible process. Most scientists in the late 19th and even early 20th century negatively related to the Maxwell-Boltzmann statistical theory, because of the obvious contradiction between the reversibility of the laws of mechanics and the irreversibility postulated by the second law of thermodynamics. Many scientists, supporters of the thermodynamic-energy worldview, denied even the existence of atoms and their perpetual thermal motion because of the belief in the second law of thermodynamics \cite{Smoluchowski}.  

The investigations of the Brownian motion by Einstein, Smoluchowski and others have convinced even supporters of the thermodynamic - energy worldview in the existence of perpetual thermal motion of atoms, molecules, electrons, ions, and other particles. Smoluchowski wrote in 1914: "{\it Atomistics is recognized as the basis of modern physics in general; the second law of thermodynamics has once and for all lost its significance as an unshakable dogma, as one of the basic principles of physics}" \cite{Smoluchowski}. But the dogma has changed rather than lost its significance. Scientists of the 19th century were rejecting any perpetual motion at thermodynamic equilibrium, whereas most modern scientists reject a possibility of any directed thermal motion at thermodynamic equilibrium. 

Almost all physicists believed in the impossibility of directed thermal motion. Only the great scientist Max Planck was understanding that this belief has no scientific basis. Therefore he questioned the Boltzmann H - theorem: "{\it Boltzmann omitted in his deduction every mention of the indispensable presupposition of the validity of his theorem - namely, the assumption of molecular disorder}" \cite{Planck}. Planck could not, more than a hundred years ago, have given examples of violation of the assumption of molecular disorder. Therefore his doubt about the H - theorem was probably considered by those who knew about this doubt as a consequence only of a pedantry of Planck. The observations \cite{Letter2007,LP1962,PC2007,Science2009PC,PRL2009PC} of the persistent current $\overline{I_{p}} \neq 0$ undamped at a non-zero resistance $\overline{R} > 0$ testify that Planck's pedantry turned out to be genius. Measurements \cite{Physica2019} of the dc voltage $V_{dc} \propto I_{p}$ and the dc power $V_{dc}I_{p}$ give experimental evidence that the energy of the persistent current can be used for an useful work.  

If physicists had paid attention to the reason for Planck's doubts about the Boltzmann H - theorem, then perhaps they would not have contradicted logic after 1933, and the authors \cite{Science2009PC,Birge2009} would not make the preposterous claim contradicting to the elementary mathematics. The belief in the second law of thermodynamics determined the entire history of the creation of the theory of superconductivity, which was created within the framework of equilibrium thermodynamics, despite Joule heating. The physical community should refuse the blind belief in the second law of thermodynamics. This belief has led to the internal inconsistency of the conventional theories of superconductivity, which Hirsch points out \cite{Hirsch2020Physica,Hirsch2020EPL,Hirsch2020ModPhys}, and contradicts the experimental results \cite{Physica2019,Letter2007,LP1962,PC2007,Science2009PC,PRL2009PC}. Now, as in the thirties of the last century, almost all physicists believe in the second law of thermodynamics. Nevertheless, some publications of articles and even book \cite{Book05} call into question absolute status of the second law of thermodynamics. The author of the articles \cite{Peter2004,Peter2010} was considering a contradiction of the Meissner effect with the second law of thermodynamics.    
          
The author is grateful to the Reviewer for his useful comments and the quote from Keezom's publication \cite{Keesom1934} used in the revised version of the article. This work was made in the framework of State Task No 075-00355-21-00.

\end{document}